\documentclass[11pt]{article}

\usepackage{amsmath,amsthm,amssymb}
\usepackage{epsfig}

\newcommand{\1}{\,\mathbb{I}}
\newcommand{\bra}[1]{\left\langle{}#1\right|}
\newcommand{\cfield}{\mathbb{C}}
\newcommand{\ctp}{C}
\newcommand{\dff}{\sc}
\newcommand{\dmn}{\mathcal{D}}
\newcommand{\ee}{\mathbf{e}}

\newcommand{\hh}{\mathcal{H}}

\newcommand{\ket}[1]{\left| #1\right\rangle}
\newcommand{\kf}{\mathbf{F}}
\newcommand{\kk}{\mathbf{K}}

\newcommand{\lth}{n}

\newcommand{\ppp}[1]{{#1}{#1'}}

\newcommand{\rrh}{\rho}

\DeclareMathOperator{\trc}{Tr}

\newcommand{\cbn}{\cfield{B}_\lth}

\newcommand{\cbp}{\mathfrak{T}}
\newcommand{\bracket}[3]{\bra{#1}#2\ket{#3}}

\newcommand{\braket}[2]{\left\langle{}#1\,\right|\left.#2\right\rangle}
\newcommand{\hhp}{\mathfrak{B}}
\newcommand{\hlp}{\mathfrak{L}}
\newcommand{\ketbra}[2]{\ket{#1}\!\!\bra{#2}}

\newcommand{\mesb}{\,d\mathbf{S}_{\lth}}

\newcommand{\rrp}{\boldsymbol{\rrh}}

\newcommand{\raypr}[1]{\ketbra{#1}{#1}}

\title{
An asymptotical separability criterion for bipartite density
operators}

\author{Rom\`an R. Zapatrin\thanks{Friedmann Lab. for Theoretical
Physics, SPb EF University, Griboyedova 30--32, 191023,
St.Petersburg, Russia; e-mail: zapatrin@rusmuseum.ru}}

\begin{document}

\maketitle

\begin{abstract}
For a given density matrix $\rho$ of a bipartite quantum system an
asymptotical separability criterion is suggested. Using the
continuous ensemble method, a sequence of separable density
matrices is built which converges to $\rho$ if and only if $\rho$
is separable. The convergence speed is evaluated and for any given
tolerance parameter $\kappa$ an  iterative procedure is suggested
which decides in finite number of steps if there exists a
separable density matrix $\rho_\kappa$ which differs from the
matrix $\rho$ by at most $\kappa$.
\end{abstract}

\section{Introduction}\label{sintro}

The notion of entanglement plays a central r\^ole in quantum
communication and quantum computation. Usually the efforts are
focused on quantifying entanglement itself, that is, describing
the \emph{impossibility} to prepare a state by means of LOCC
(local operations and classical communications). One may,
although, try to quantify \emph{separability} rather than
entanglement, this was shown to be a tool for geometrical
classification of mixed multipartite entangled states
\cite{myjmo}. I dwell on the case of bipartite quantum systems. A
state of such system is called separable if it can be prepared by
LOCC. In terms of density matrices that means that $\rrp$, its
density matrix, can be represented as a mixture of pure product
states. I suggest to replace finite sums of projectors by
continuous distributions on the set of unit vectors.

Recently continuous ensemble method was introduced \cite{mygibbs}
which was applied to explore the separability of
finite-dimensional quantum systems \cite{coei,coeii}. First, a
geometrical characterization of robustly separable mixed states
was provided in \cite{coei}, then the separability problem was
reduced to a finite number of numerical equations \cite{coeii}.
This paper suggests an iterative procedure which, for any given
tolerance parameter, decides if the state is separable or not for
finite number of steps.

This is done in the following way. In bipartite case the density
operators are represented by distribution on the Cartesian product
of unit spheres in subsystems' spaces. Given a density operator
$\rrp$, we consider it as an element of the space $\hlp$ of all
self adjoint operators in the product space $\hhp=\hh\otimes\hh$.
The density operator $\rrp$ is called {\dff robustly separable} if
it has a neighborhood of separable operators \cite{vidaltarrach}.
The robust separability of $\rrp$ is shown \cite{coeii} to be
equivalent to the solvability of the following vector equation in
$\hlp$:

\[
\nabla
\kk
\;=\;
\rrp
\]

\noindent with respect to the trace functional $\kk$ on $\hlp$
introduced in the first paper \cite{coei} of this series. When we
fix a product basis in $\hhp$, this equation becomes a system of
$\lth^4$ transcendent equations with respect to $\lth^4$ variables
\cite{coeii}.

In this paper an asymptotic procedure based on iterated function
method is suggested which, in case if $\rrp$ is separable,
approximates $\rrp$ by a sequence of separable density operators,
otherwise this sequence diverges. The operator $\kf$ is introduced

\[
\kf(X)
\;=\;
X + \lambda
\left( \,
\rrp -
\iint
e^{\bracket{\ppp{\phi}}{X}{\ppp{\phi}}}
\raypr{\ppp{\phi}}
\ppp{\mesb}
\right)
\]

\noindent where the integration is taken over the torus---the
Cartesian product of unit spheres in $\hh,\hh'$. Then the sequence
of operators $X_0,X_1,\ldots,X_m,\ldots$ is built

\[
\left\lbrace
\begin{array}{l}
  X_0=0 \\
  X_1=\kf(X_0) \\
  \ldots \\
  X_m=\kf(X_{m-1})  \\
  \ldots
\end{array}
\right.
\]

\noindent which converges if $\rrp$ is robustly separable. The
constant $\lambda$ and the speed of this convergence are (roughly)
evaluated, they depend on the dimension $\lth$ of the state space
and the tolerance $\kappa$ we choose.

\section{Continuous optimal ensembles}\label{scontrem}

In this section I briefly summarize the basics of continuous
ensembles approach to quantifying quantum separability. In general
context related to coherent states, continuous ensembles were
considered yet by Glauber \cite{glauber} and Sudarshan
\cite{sudarshan}. Generalizing the fact that any convex
combination of density operators is again a density operator, we
represent density operators as probability distributions on the
unit sphere\footnote{Pure states form a projective space rather
than the unit sphere in $\hh$. On the other hand, one may
integrate over any probabilistic space. For technical reasons I
prefer to represent ensembles of pure states by measures on unit
vectors in $\hh$. I use the Umegaki measure on $\cbn$--- the
uniform measure with respect to the action of $U(\lth)$ normalized
so that $\int_{\cbn}\mesb=1$. Similarly, for bipartite case the
integration will be carried out over the Cartesian product of unit
spheres in appropriate state spaces.} in the state space $\hh$ of
the system (these ideas were also considered in \cite{sqprop}). In
particular, the well-known spectral decomposition of an operator
is a special case of such representation, this will be addressed
below. However the main feature of the suggested approach is that
the distribution is taken over \emph{non-orthogonal} states.

\medskip

Begin with the case of a single quantum system. Consider a
probability distribution on $\cbn$ whose density is a function
$\mu(\phi)$, where $\phi\in\cbn$ ranges over all unit vectors in
$\hh$. The density operator of this continuous ensemble is:

\begin{equation}\label{e01integral}
  \rrh
  \;=\;
  \int_{\phi\in\cbn}\limits\;
  \mu(\phi)\,
  \raypr{\phi}
  \,\mesb
\end{equation}

\noindent where $\raypr{\phi}$ is the projector onto the vector
$\bra{\phi}$ and $\mesb$ is the normalized measure on $\cbn$.
Effectively the operator integral $\rrh$ \eqref{e01integral} can
be calculated by its matrix elements. In any fixed basis
$\{\ket{\ee_i}\}$ in $\hh$, the matrix elements
$\rrh_{ij}=\bracket{\ee_i}{\rrh}{\ee_j}$ are the following
\emph{numerical} integrals:

\[
\rrh_{ij}
\;=\;
  \bracket{\ee_i}{\rrh}{\ee_j}
  \;=\;
  \int_{\phi\in\cbn}\limits\;
  \mu(\phi)\,
  \braket{\ee_i}{\phi}
  \braket{\phi}{\ee_j}
  \,\mesb
\]

\paragraph{Smeared spectral decomposition.} Since we are
interested in robustly separable states, let us restrict ourselves
to full-range density operators. The usual spectral decomposition
$\rrh=\sum{}p_k\raypr{\ee_k}$ can be treated as an atomic measure
on $\cbn$ whose density is the appropriate combinations of delta
functions:

\[\mu_{\mbox{\scriptsize spec}}(\phi)=\sum{}p_k\,\delta({\phi-\ee_k})\]

We introduce a `smeared' version of the spectral decomposition.
Let $K$ be an integer number, then the density matrix $\rrh$ is
represented as a continuous ensemble:

\[
\rrh
\;=\;
\sum{}p_k\raypr{\ee_k}
\;=\;
  \int_{\phi\in\cbn}\limits\;
\mu(\phi)
\,
 \raypr{\phi}
  \,\mesb
\]

\noindent The density $\mu$ was calculated in \cite{coeii}, it
equals

\begin{equation}\label{esmsing}
\mu(\phi)
\;=\;
\frac{((K+1)\lth)!}{K\cdot(K\lth)!
\,\lth!}
\,\cdot
\sum_k\limits
\left(p_k
\,-
\frac{1}{(K+1)\lth}
\right)
\left|
\braket{\ee_k}{\phi}
\right|^{\,2Kn}
\end{equation}

\noindent Furthermore, the distribution \eqref{esmsing} tends to
the spectral decomposition of $\rrh$ as $K$ tends to infinity.
Denote by $p_0$ the smallest eigenvalue of the density matrix
$\rrh$ (recall that $p_0>0$ as we restrict ourselves to full-range
density matrices). Then the density $\mu$ in \eqref{esmsing} is
positive for any such $K$ that $(K+1)\lth\ge 1/p_0$. We may take

\begin{equation}\label{edefksingle}
  K
\;=\;
\frac{1}{\lth\,p_0}
\end{equation}

\paragraph{Optimal entropy ensembles.} It was shown in
\cite{coei,coeii} that, given a robustly separable density matrix
$\rrp$ in the product space $\hhp=\hh\otimes\hh'$, there exists a
self-adjoint operator $X$ in $\hhp$ such that $\rrp$ can be
represented as the following operator integral:

\begin{equation}\label{erepbi}
\rrp
\;=\;
\iint_{\ppp\phi\in\cbp}\limits\;
\;e^{\bracket{\ppp\phi}{X}{\ppp\phi}}
\raypr{\ppp\phi}\ppp\mesb
\end{equation}

\noindent Fix a product basis $\{\bra{\ee_i{\ee'}_{i'}}\}$. Then
the condition \eqref{erepbi} takes the form

\begin{equation}\label{erepbas}
\bracket{\ee_i{\ee'}_{i'}}{\rrp}{\ee_j{\ee'}_{j'}}
\;=\;
\iint_{\ppp\phi\in\cbp}\limits
\;e^{\bracket{\ppp\phi}{X}{\ppp\phi}}
\braket{\ee_i{\ee'}_{i'}}{\ppp\phi}
\braket{\ppp\phi}{\ee_j{\ee'}_{j'}}
\ppp\mesb
\end{equation}

\noindent and the separability problem reduces to finding the
operator $X$ from this system of equations. The essence of the
separability problem is the \emph{existence} of a solution of
\eqref{erepbas}. In this paper I suggest an asymptotic procedure
which gives the answer up to an arbitrary (but finite) precision.

\section{Contraction mappings and
the iterated function method}\label{scontrmap}

A {\dff contraction mapping} is any mapping $\kf:\dmn\to\dmn$ in a
metric space $\dmn$ such that the distance $d$ between any two
points before the mapping is greater than the distance between
them after the mapping:

\begin{equation}\label{edefcontr}
\exists
\;\ctp<1\quad
\forall
\;x,y\in\dmn
\qquad
d(\kf(x),\kf(y))
\;\le
\;\ctp\cdot\,d(x,y)
\end{equation}

The most important property of a contraction mapping is that there
is exactly one point which is invariant under the mapping. The
second crucial feature is that under sufficiently long iteration
every point eventually contracts to this point. Using the
definition \eqref{edefcontr}, the speed of convergence can be
evaluated from the value of the contraction parameter $\ctp$ and
the diameter $D$ of the space $\dmn$. After $N$ steps the distance
between the fixed point of $\kf$ and its $N$-th approximation will
not exceed $\ctp^N D$, therefore the given accuracy $\kappa$ will
be achieved in at most

\begin{equation}\label{ensteps}
N
\;=\;
\frac{\ln\kappa-\ln D}{\ln\ctp}
\end{equation}

\noindent steps. A standard way to solve the equation $f(x)=a$
using the iterated function method is to define the function
$F(x)$:

\[
F(x)
\;:=\;
x+\lambda(a-f(x))
\]

\noindent and choose a domain $\dmn$ for $x$ such that $f$ acts
within $\dmn$ and a parameter $\lambda$ in order to make $f$
contracting on $\dmn$. If the derivative $f'$ is bounded on
$\dmn$, then the contraction condition can always be satisfied by
choosing sufficiently small $\lambda$, namely, take any

\[
  \lambda
\,<\,
\frac{1}{\;\max|f'|}
\]

\noindent In this case the contraction parameter in
\eqref{edefcontr} is $\ctp=\lambda/\max|f'|$.

\bigskip

Now let us apply this method to the equation \eqref{erepbas}.
Introduce the mapping $F:\hlp\to\hlp$:

\begin{equation}\label{edefcf}
\kf(X)
\;=\;
X+
\lambda\,
\left(\rrp\;-
  \iint_{\ppp\phi\in\cbp}\limits\;
  e^{\bracket{\ppp\phi}{X}{\ppp\phi}}
\,\raypr{\ppp\phi}\ppp\mesb
\right)
\end{equation}

\noindent with the domain $\dmn=\{X\,:\,\left\|X\right\|\le A\}$
for a sufficiently large $A$, see section \ref{sevaliter} for
evaluations of the parameters in \eqref{edefcf}. If such $X$
exists, then

\begin{itemize}
  \item we can always choose $\lambda$ which makes $\kf$ contraction
mapping (this is because $e^{\bracket{\ppp\phi}{X}{\ppp\phi}}$ is
bounded on $\dmn$)
  \item whatever be $\rrp$, the zero operator 0 is always in $\dmn$
\end{itemize}

This opens us the way to solve the separability problem. Start the
iteration procedure with the zero operator, then, if the solution
exists, the sequence will converge, otherwise not.

\section{Evaluating the parameters}\label{sevaliter}

In order to evaluate the parameters for the applicability of the
iterated function method, we need to specify
\begin{itemize}
  \item The domain $\dmn$---to make the function $\kf$ act within
$\dmn$
  \item The value of the parameter $\lambda$---to make the mapping $\kf$
contracting
\end{itemize}

\medskip

Assume that the density operator $\rrp$ is separable, then there
exists a finite set of 1-dimensional product projectors
$\{\bra{\mathbf{f}_\alpha\mathbf{f'}_\alpha}\}$ such that $\rrp$
is a convex combination

\[
\rrp
\;=\;
\sum_\alpha
\,c_\alpha
\raypr{\mathbf{f}_\alpha
\mathbf{f'}_\alpha}
\]

\noindent The (real) dimension of the state of all self-adjoint
operators in the product space $\hhp$ is $\lth^4$. According to
Karath\'eodori theorem, the projectors
$\raypr{\mathbf{f}_\alpha\mathbf{f'}_\alpha}$ can be chosen so
that their number in the above sum will not exceed $\lth^8$.

\medskip

Now recall that we have a tolerance parameter $\kappa$. What we
need, is to assume that each product projector
$\raypr{\mathbf{f}_\alpha\mathbf{f'}_\alpha}$ can be replaced by
the mixture

\begin{equation}\label{errpalpha}
\rrp_\alpha
\;=\;
\left(
\vphantom{\sum_a^a\limits}
\left(1-
\frac{\kappa}{\lth^8}
\right)
\,\cdot
\raypr{\mathbf{f}_\alpha}
\,+\,
\frac{\kappa}{\lth^8}\1
\right)
\otimes
\left(
\vphantom{\sum_a^a\limits}
\left(1-
\frac{\kappa}{\lth^8}
\right)
\,\cdot
\raypr{\mathbf{f'}_\alpha}
\,+\,
\frac{\kappa}{\lth^8}\1
\right)
\end{equation}

\paragraph{Smeared spectral decomposition for the summands.} Each
summand in \eqref{errpalpha} is a product of density matrices in
single-party spaces. Represent it as a continuous mixture

\begin{equation}\label{errpa}
\rrp_\alpha
\;=\;
\left(
\;\int_{\cbn}\limits\;
\mu_\alpha(\phi)\,
\raypr{\phi}\mesb
\right)
\otimes
\left(
\;\int_{\cbn}\limits\;
\mu_\alpha(\phi')\,
\raypr{\phi'}\mesb'
\right)
\end{equation}

\noindent Let us first evaluate the range $\mu_0<\mu<\mu_1$ of the
density $\mu(\phi)$ for the single particle case. First note that
the values of the uniform function $f=\sum{}c_k{}r_k^{2m}$ on the
sphere  $\sum{}r_k^{2}=1$ ranges between

\[
\frac{c_0}{\lth^{m-1}}
\le{}f(r_1,\ldots,r_\lth)\le{}c_1
\]

\noindent where $c_0$ and $c_1$ are the least and the greatest
coefficients, respectively. Then it follows from \eqref{esmsing}
that

\[
\frac{((K+1)\lth)!}{K\cdot(K\lth)!
\,\lth!}
\,\cdot\,
\frac{1}{\lth^{K\lth-1}}
\,
\left(p_0
\,-
\frac{1}{(K+1)n}
\right)
\le\mu\le
\frac{((K+1)\lth)!}{K\cdot(K\lth)!
\,\lth!}
\,\cdot
\left(p_1
\,-
\frac{1}{(K+1)n}
\right)
\]

\noindent Denote

\begin{equation}\label{edefck}
C_K
\;=\;
\frac{((K+1)\lth)!}{K\cdot(K\lth)!
\,\lth!}
\end{equation}

\noindent For evaluation purposes replace $p_1-\frac{1}{(K+1)n}$
by $1$. Substituting $K=1/\lth p_0$ from \eqref{edefksingle} and
calculating $p_0-\frac{1}{(K+1)n}=\frac{1}{\lth}\cdot
\frac{\lth^2
p_0^2}{1+\lth{}p_0}>\frac{1}{\lth}\cdot
\frac{\lth^2
p_0^2}{2}$ we obtain

\begin{equation}\label{eckmu}
C_K\,\cdot\,
\frac{1}{\lth^{K\lth}}
\,\cdot\,
\frac{\lth^2 p_0^2}{2}
\le\mu\le
\,C_K
\end{equation}

\noindent Recall that we consider our density operator up to a
tolerance parameter $\kappa$, hence we may assume from
\eqref{errpalpha} that the smallest eigenvalue $p_0$ in
\eqref{esmsing} to be at least $\kappa/\lth^8$, therefore $K$ can
be evaluated from \eqref{edefksingle} as

\begin{equation}\label{edefksinlev}
K
\;=\;
\frac{1}{\lth p_0}
\;=\;
\frac{\lth^7}{\kappa}
\end{equation}

\noindent Continue the evaluation \eqref{eckmu} with
$p_0=\kappa/\lth^8$:

\[
C_K\,\cdot\,
\frac{1}{\lth^{(\lth^8/\kappa)}}
\,\cdot\,
\frac{\kappa^2}{2\lth^{14}}
\le\mu\le
\,C_K
\]

\noindent Then denote

\begin{equation}\label{edefca}
C_A
\;=\;
\frac{2\lth^{14}\lth^{(\lth^8/\kappa)}}{\kappa^2}
\end{equation}

\noindent Taking into account that $C_K, C_A>>1$, make the
evaluation of $\mu$ rough but symmetric:

\[
\frac{1}{C_K\cdot C_A}
\le\mu\le
\,C_K\cdot C_A
\]

\noindent then the magnitude of the density representing of each
product $\rrp_\alpha$ in \eqref{errpa} ranges within

\[
\left(
\frac{1}{C_K C_A}
\right)^2
\le\mu_\alpha(\phi)\cdot\mu_\alpha(\phi')\le
\,C_K^2\,
C_A^2
\]

\noindent Since the initial density matrix $\rrp$ is a convex
combinations of the matrices $\rrp_\alpha$, the matrix $\rrp$
itself can be represented as a continuous ensemble whose density
ranges within the same bounds \eqref{eboundsmu}. That means that
the values of the operator $X$ satisfy the following relations

\begin{equation}\label{eboundsmu}
\left(
\frac{1}{C_K C_A}
\right)^2
\,\le\,
e^{\bracket{\ppp\phi}{X}{\ppp\phi}}
\,\le\,
\,C_K^2
\,C_A^2
\end{equation}

\noindent This gives us the domain for the function $\kf$ we are
going to iterate:

\begin{equation}\label{edmnx}
\lVert
X
\rVert
\,\le\,
2\ln
\left(
{C_K C_A}
\right)
\end{equation}

\noindent where the expression for $C_K$ and $C_A$ are obtained
above in \eqref{edefck} and \eqref{edefca}, respectively, and the
norm is $L_1$.

\paragraph{Evaluating the derivative.} In order to learn the value
of the parameter $\lambda$ in \eqref{edefcf} the derivative of
$\kf$ in any arbitrary direction must be evaluated. Let $Y$ be a
self-adjoint operator in $\hhp$, consider the operator $\kf(X+tY)$
and calculate its derivative in the direction $Y$ at $t=0$:

\begin{equation}\label{edefd}
D
\;:=\;
\left.
\frac{d}{dt}
\kf(X+tY)
\right|_{t=0}
\;=\;
Y-
\lambda\,
  \iint_{\ppp\phi\in\cbp}\limits\;
  e^{\bracket{\ppp\phi}{X}{\ppp\phi}}
\,{\bracket{\ppp\phi}{Y}{\ppp\phi}}
\,\raypr{\ppp\phi}\ppp\mesb
\end{equation}

\noindent Note that, according to \eqref{eboundsmu}, the values of
$e^{\bracket{\ppp\phi}{X}{\ppp\phi}}$ are bounded by $(C_K
C_A)^2$, therefore, in order to evaluate the derivative, we take
the maximal value, so the value of $D$ is bounded

\[
\lVert
D
\rVert
\;\le\;
\left\lVert\,
Y-
\lambda\,
\left(C_K C_A\right)^2
  \iint_{\ppp\phi\in\cbp}\limits\;
\,{\bracket{\ppp\phi}{Y}{\ppp\phi}}
\,\raypr{\ppp\phi}\ppp\mesb
\right\rVert
\]

\medskip

In order to evaluate the integral I use the formula obtained in
\cite{mygibbs} for the single particle case. For any self-adjoint
operator $A$ in $\hh$

\[
  \int_{\phi\in\cbn}\limits\;
\bracket{\phi}{A}{\phi}
 \ketbra{\phi}{\phi}
  \,\mesb
\;=\;
\frac{A+\1\cdot\trc A}{\,\lth(\lth+1)}
\]

\noindent Recall that the $L_1$ norm is used, this makes it
possible to evaluate the traces of $Y$ as they are bounded by the
norm of $Y$, then

\begin{equation}\label{enorms}
\left\lVert
\;A
\;-
\int_{\phi\in\cbn}\limits\;
\bracket{\phi}{A}{\phi}
 \ketbra{\phi}{\phi}
  \,\mesb
\right\rVert
\;\le\;
\|A\|
\cdot
\left(
1-\frac{1}{\lth(\lth+1)}
\right)
\end{equation}

\noindent Since the integral \eqref{edefd} is a linear function of
the operator $Y$,

\[
  \iint_{\ppp\phi\in\cbp}\limits\;
\,{\bracket{\ppp\phi}{Y}{\ppp\phi}}
\,\raypr{\ppp\phi}\ppp\mesb
\;=\;
\frac{Y+\1\otimes\trc_{\mbox{\scriptsize{}I}}{}Y+
\trc_{\mbox{\scriptsize{}II}}{}Y\otimes\1+
\trc{}Y\cdot\1\otimes\1}{\,\lth^2\,(\lth+1)^2}
\]

\noindent where
$\trc_{\mbox{\scriptsize{}I}},\trc_{\mbox{\scriptsize{}II}}$ stand
for partial traces. So, taking into account \eqref{enorms}

\[
\lVert
D
\rVert
\;\le\;
\left\lVert\,
Y-
\lambda\,
\left(
C_K C_A
\right)^2
\frac{Y+\1\otimes\trc_{\mbox{\scriptsize{}I}}{}Y+
\trc_{\mbox{\scriptsize{}II}}{}Y\otimes\1+
\trc{}Y\cdot\1\otimes\1}{\,\lth^2\,(\lth+1)^2}
\right\rVert
\]

\noindent When we set $\lambda$ to be

\begin{equation}\label{evlambdanew}
  \lambda
\;=\;
\frac{1}{\left(
C_K C_A\right)^2}
\end{equation}

\noindent we have the following evaluation of contraction constant
$\|D\|\le\ctp\|Y\|$ in \eqref{edefcontr}:

\begin{equation}\label{enormb}
\ctp
\;=\;
\left(
1-\frac{1}{\lth(\lth+1)}
\right)^2
\end{equation}

\paragraph{Evaluating the speed of convergence.} According to
\eqref{edmnx} we have the following evaluation of the diameter of
the domain $X$

\[
D
\;\le\;
4\ln(C_A{}C_K)
\]

\noindent Then, fixing a tolerance parameter $\kappa$ and using
the calculated contraction constant \eqref{enormb} we can apply
\eqref{ensteps} to find the sufficient number of iterations

\[
N
\;=\;
\frac{\ln\kappa-\ln\bigl(4\ln(C_A{}C_K)\bigr)}{2
\ln\left(1-\frac{1}{\lth(\lth+1)}\right)}
\]

\noindent and approximating the logarithm in the denominator we
get

\begin{equation}\label{edefn}
N
\;=\;
2\lth(\lth+1)
\left(
\ln\bigl(4\ln(C_A{}C_K)\bigr)+\ln\frac{1}{\kappa}
\right)
\end{equation}

\section{Concluding remarks}\label{sconclud}

Given a density matrix $\rrp$ in $\hhp$, a question arises if it
is separable or not. When the dimension of at least one of spaces
$\hh,\hh'$ is 2, this question was given an effective and exact
answer---the positive partial transpose (PPT) criterion due to
Peres-Horodecki was suggested \cite{peres,horod}. The criterion
states that $\rrp$ is separable if and only if its partial
transpose $\rrp^{T_{II}}$ remains non-negative matrix. In higher
dimensions PPT is only a necessary condition for a state to be
factorizable as there exist entangled density matrices whose
partial transpose if positive.

In this paper the dimension of $\hh,\hh'$ may take any finite
value $\lth$. The separabilty problem is resolved with any given
in advance finite precision. Let us summarize the procedure.
\begin{itemize}
  \item Given a density operator $\rrp$ in the product space
$\hhp=\hh\otimes\hh'\simeq\cfield^{\lth^2}$, calculate its minimal
eigenvalue $p_0$.
  \item Set up the tolerance parameter $\kappa$ and calculate the
values of $K$ \eqref{edefksinlev}, $C_K$ \eqref{edefck} and $C_A$
\eqref{edefca}:
\[
K
\;=\;
\frac{\lth^7}{\kappa}
\quad;\qquad
C_K
\;=\;
\frac{((K+1)\lth)!}{K\cdot(K\lth)!
\,\lth!}
\quad;\qquad
C_A
\;=\;
\frac{2\lth^{14}\lth^{(\lth^8/\kappa)}}{\kappa^2}
\]
\noindent from which we derive the value of $\lambda$
\eqref{evlambdanew} and the number of steps $N$ \eqref{edefn}:
\[
  \lambda
\;=\;
\frac{1}{\left(
C_K C_A\right)^2}
\quad;\qquad
N
\;=\;
2\lth(\lth+1)
\left(
\ln\bigl(4\ln(C_A{}C_K)\bigr)+\ln\frac{1}{\kappa}
\right)
\]
  \item Iterate the function $\kf$ \eqref{edefcf}
\[
X_{k+1}
\;=\;
\kf(X_k)
\;=\;
X_k+
\lambda\,
\left(\rrp\;-
  \iint_{\ppp\phi\in\cbp}\limits\;
  e^{\bracket{\ppp\phi}{X_k}{\ppp\phi}}
\,\raypr{\ppp\phi}\ppp\mesb
\right)
\]
\noindent starting from $X_0=0$. At each step check $||X||_1$ and
if it occurs that $||X||_1>2\ln\left( C_K C_A\right)$ that signals
us that $\rrp$ is entangled.
  \item After $N$ steps we get the approximation of $\rrp$ with
accuracy $\kappa$
\end{itemize}

\medskip

I emphasize that this method is not completely suited for
practical calculations as the evaluations for the convergence
parameters are very rough. This paper just demonstrates the
principal possibility of using the continuous ensemble method to
tackle the separability problem.

\paragraph{Acknowledgments.} I would like to express my gratitude
to many people who offered me their attention for a pretty long
time while this work was carried out. These are Abhay Ashtekar,
Paolo Giorda, Markus Grassl, Radu Ionicioiu, Serguei Krasnikov,
Rafael de la Madrid, Ioannis Raptis, Mario Rasetti, Martin
R\"otteler, Ujjwal Sen, Karl Svozil, Xiaoguang Wang, Pawe\l{}
Wojcian, Christof Zalka and, alphabetically the last but
definitely not the least, Paolo Zanardi.

The financial support for the initial stage of this research was
provided by the EU FP5 Project Q-ACTA (in the I.S.I. Foundation,
Torino, Italy). Crucial issues related to this research were
intensively discussed during the meeting Glafka-2004 `Iconoclastic
Approaches to Quantum Gravity' (15--18 June, 2004, Athens, Greece)
supported by QUALCO Technologies (special thanks to its
organizers---Ioannis Raptis and Orestis Tsakalotos). The financial
support on the final stage was provided by the research grant No.
04-06-80215a from RFFI (Russian Basic Research Foundation).


\begin{thebibliography}{99}

\bibitem{glauber}
R. J. Glauber, \emph{Coherent and Incoherent States of the
Radiation Field}, Phys. Rev. \textbf{131}, 2766--2788  (1963)

\bibitem{horod}
P. Horodecki, \emph{Separability criterion and inseparable mixed
states with positive partial transposition}, Phys. Lett. \textbf{A
232}, 333 (1997); eprint quant-ph/9703004

\bibitem{sqprop}
E. Lehrer, E. Shmaya, \emph{A Subjective Approach to Quantum
Probability}; eprint quant-ph/0503066

\bibitem{peres}
A. Peres, \emph{Separability Criterion for Density Matrices},
Phys. Rev. Lett. \textbf{77}, 1413 (1996); eprint quant-ph/9604005

\bibitem{sudarshan}
E. C. G. Sudarshan, \emph{Equivalence of Semiclassical and Quantum
Mechanical Descriptions of Statistical Light Beams}, Phys. Rev.
Lett. \textbf{10}, 277--279 (1963)

\bibitem{vidaltarrach}
G. Vidal, R. Tarrach, \emph{Robustness of entanglement}, Phys.
Rev. \textbf{A59}, 141--155 (1999); eprint quant-ph/9806094

\bibitem{myjmo}
R.R.Zapatrin, \emph{Combinatorial Topology Of Multipartite
Entangled States}, Journal of Modern Optics, \textbf{50}, 891--899
(2003); eprint quant-ph/0207058

\bibitem{mygibbs}
R.R.Zapatrin, \emph{A note on continuous ensemble expansions of
quantum states}; eprint quant-ph/0403105

\bibitem{coei}
R.R.Zapatrin, \emph{Continuous optimal ensembles I: A geometrical
characterization of robustly separable quantum states}; eprint
quant-ph/0503173

\bibitem{coeii}
R.R.Zapatrin, \emph{Continuous optimal ensembles II. Reducing the
separability condition to numerical equations}; eprint
quant-ph/0504034

\end{thebibliography}
\end{document}